\documentclass[usenatbib]{mn2e} 
\usepackage[mathscr]{eucal}
\usepackage{pslatex,amsmath,amsfonts,amssymb,amscd}
\usepackage{graphicx}
\usepackage{aas_macros}
\newcommand\bl{{\mathbf l}}

\newcommand\br{{\mathbf r}}

\newcommand\bx{{\mathbf x}}

\newcommand\by{{\mathbf y}} 
\newcommand\bz{{\mathbf z}}
\newcommand\bzero{{\mathbf 0}}

\newcommand\bA{{\mathbf A}} 
\newcommand\bB{{\mathbf B}}

\newcommand\bE{{\mathbf E}} 
 
\newcommand\bG{{\mathbf G}} 
 
\newcommand\bI{{\mathbf I}} 
\newcommand\bJ{{\mathbf J}} 
 
\newcommand\bL{{\mathbf L}} 
 
\newcommand\bN{{\mathbf N}}
  
\newcommand\bP{{\mathbf P}}
 
\newcommand\bR{{\mathbf R}}

\newcommand\bX{{\mathbf X}} 
\newcommand\bY{{\mathbf Y}} 
\newcommand\bZ{{\mathbf Z}} 
\newcommand\scB{{\mathscr B}}

\newcommand\scP{{\mathscr P}}

\def\mvector#1{{\pmb #1}}

\newcommand\vc{\mvector{c}}

\newcommand\vl{\mvector{l}}

\newcommand\vr{\mvector{r}}

\newcommand\vx{\mvector{x}}
\newcommand\vy{\mvector{y}}
\newcommand\vz{\mvector{z}}

\newcommand\vX{\mvector{X}}
\newcommand\vY{\mvector{Y}}
\newcommand\vZ{\mvector{Z}}
\newcommand\field{\mathbb}

\newcommand\R{\field{R}}

\newcommand\bOmega{\boldsymbol{\Omega}}

\newcommand\diag{\operatorname{diag}}

\newcommand\Dt{\frac{\mathrm{d}\phantom{t} }{\mathrm{d}\mspace{1mu} t}}

\newcommand\pder[2]{\dfrac{\partial #1 }{\partial #2}} 
\newcommand\mtext[1]{\quad\text{#1}\quad}

\newcommand\defset[2]{\left\{{#1}\;\vert \;\; {#2} \,\right\}}

\newcommand\pairing[2]{\langle {#1},{#2} \rangle}

\newcommand\ed[1]{{#1}}

\usepackage{color}  
\definecolor{myred}{rgb}{0.7,0.0,0.2}
\newtheorem{theorem}{Theorem}[section]
\newtheorem{proposition}[theorem]{Proposition}
\title[The two-body problem of Kinoshita revisited]%
{Relative equilibria in the unrestricted problem of a sphere and symmetric rigid body}
\author[M. Vereshchagin, A. J. Maciejewski \&  K. Go\'zdziewski]{
Mikhail Vereshchagin$^{1}$\thanks{e-mail: m.vereshchagin@gmail.com},
Andrzej J. Maciejewski$^{2}$\thanks{e-mail: maciejka@astro.ia.uz.zgora.pl} and
Krzysztof Go\'zdziewski$^{1}$\thanks{e-mail: k.gozdziewski@astri.umk.pl}\\
$^{1}$Toru\'n Centre for Astronomy, Nicolaus Copernicus University, 
Gagarin Str. 11, 87-100 Toru\'n, Poland \\
$^{2}$Institute of Astronomy, University of Zielona G\'ora,
Podg\'orna 50, PL-65--246 Zielona G\'ora, Poland
}
\begin{document}
%_______________________________________________________________________________
\date{Accepted 2009 December 3.  Received 2009 December 2; in original form 2009 September 10}
\pagerange{\pageref{firstpage}--\pageref{lastpage}} \pubyear{2009}
\maketitle
\label{firstpage}
%_______________________________________________________________________________
\begin{abstract}
We consider the unrestricted problem of two mutually attracting rigid bodies, an
uniform sphere (or a point mass) and an axially symmetric body. We present a
global, geometric approach for finding all relative equilibria (stationary
solutions) in this model, which was already studied by Kinoshita (1970). We
extend and generalize his results, showing that the equilibria solutions may be
found by solving at most two non-linear, algebraic equations, assuming that the
potential function of the symmetric rigid body is known explicitly.  We
demonstrate that there are three classes of  the relative equilibria, which we
call {\em cylindrical}, {\em inclined co-planar}, and {\em conic} precessions,
respectively.  Moreover, we also show that in the case of conic precession,
although the relative orbit is circular, the point-mass and the mass center of
the body move in different parallel planes. This solution has been yet not known
in the literature.
\end{abstract}
%_______________________________________________________________________________
\section{Introduction}
In this paper, we consider the dynamical problem of two rigid bodies interacting
mutually according to the Newtonian laws of gravitation.  It is well known that
if both these bodies are approximated by point-masses, their dynamics are
analytically solvable in terms of the integrable Kepler problem. However, even
if only one of the bodies is an arbitrary rigid body then the complexity of the
model changes dramatically. First of all, we have three more degrees of freedom
required to describe the rotational motion of the rigid body. Moreover,  because
the gravitational potential of a rigid body,  in  general case, depends on an
infinite number of parameters, the two rigid-body dynamics depend also  \ed{on
this infinite set of parameters}. Finally, orbital and rotational degrees of
freedom are coupled non-linearly, and the resulting system  is not
\ed{analytically solvable}.

Because of the mathematical complexity involved, many approximate models were studied in the
past.  For example, under certain assumptions, one can consider the co-planar
two rigid-body problem, see e.g.,
%% citation order has changed
\cite{1975MIzMU.........B,1985KosIs..23...26B,1990KosIs..28..664B,2003CeMDA..85...79G}.
Another way to simplify the problem relies on truncating series expansions of  
the gravitational potential
of the rigid body (the potential of mutual interactions), see
e.g., \cite{1967CosRe...5..457M,1967CosRe...5..318M,1985KosIs..23..323M,1988SvAL...14..118M,1989KosIs..27...31S}.
In the so-called satellite (or restricted) models, it is assumed that the rotational
motion does not influence the relative, Keplerian orbit
of the mass centres of the bodies. In such models, the relative orbit
is given parametrically, and only the rotational dynamics
\ed{need to be} investigated; for details, see
e.g.,
\cite{1985KosIs..23..323M,1967CosRe...5..457M,1975CosRe..13..285M,1994AcA....44..301M,1995CeMDA..61..347M}.

In this work, we focus on  particular version of the model \ed{involving} a
sphere (a point-mass), by considering that the rigid body is {\em axially
symmetric}. Obviously, \ed{this} leads to  \ed{much} simplified form of the
gravitational potential.  \ed{However}, in general, it still depends on an
infinite number of parameters. For the first time this problem was investigated
by \cite{1970PASJ...22..383K},  and hence it will be called {\em the Kinoshita
problem} from hereafter. Our aim is to determine all possible classes of the
relative equilibria (stationary solutions) in this problem, using a general
formalism in \cite{MR1130692} and \cite{Maciejewski:95::d}. Keeping in mind a
motivation of the work by Kinoshita, we try to generalize his analysis and to
simplify conditions for the existence of these equilibria,  %% rewording to
avoid repetition of "of" \ed{assuming only the necessary general form of the
gravitational  potential of the rigid body}. In this sense,  our paper is also
related to the work by \cite{2006CeMDA..94..317S}, who searched for the relative
equilibria in the unrestricted two rigid-body problem, and formulated conditions
of the equilibria in terms of solutions to two non-linear equations,
parameterized by integrals of the total energy and angular momentum. Moreover,
we stress that our analysis rely on the very basic vectorial form of the
equations of motion, and we consider the full, non-restricted model with
\ed{{\em symmetric}} rigid body.  The assumption of symmetry implies that the
results derived for the general, full two-body model cannot be simply
``translated'' as a particular case, and, in fact, a special reduction of the
system which takes into account the symmetry explicitly is necessary. We note
that the relative equilibria of the model with axially symmetric body may be
understood as particular {\em periodic orbits}  in the full two rigid body model
considered by \ed{\cite{2006CeMDA..94..317S}}. This reduction is in fact crucial
for a generalization of the results of Kinoshita and {\em to discover a class of
equilibria that have been missed in his analysis}. 

Investigations of the unrestricted two rigid body problem, in a version
considered in this work, and its particular stationary solutions, are
interesting because they concern a special case of generally unsolved, classic
problem of the dynamics. Moreover, the qualitative analysis of this model may
help us to answer for important, even  ``practical'' questions on the dynamics
and coupling of the rotational and orbital motions  in many astronomical
systems. These are, for instance, binary asteroids, see
\cite{2006CeMDA..94..317S} and references therein.  A deep understanding of the
qualitative dynamics are important for \ed{the} attitude determination and
control of large artificial satellites orbiting planets and/or irregular natural
moons. Recently, the diversity of extrasolar planets discovered in wide
dynamical environments also rises questions on their rotational motion and
related long term effects \ed{\cite[e.g.,][]{Correia2008}}. In the context of
mathematical complexity of the problem, the relative equilibria are the simplest
class of solutions that may be found and investigated analytically, and are
helpful to construct local, precise analytic theories of motion in their
vicinity in the phase space \ed{\cite[e.g.,][]{Gozdziewski1998}}.

A plan of this paper is the following. In Sect.~2 we formulate the mathematical
model and the equations of motion  in the most general, vectorial form are
derived. Section~3 is devoted \ed{to define the Kinoshita problem}. Next, we
introduce the relative equilibria, and we perform \ed{a global analysis} of
their existence and bifurcations (Sect.~4). This part relies on particular,
geometric reduction of the equations of motion, and is the primary key for our
analysis. Finally, we compare \ed{the classic results} by Kinoshita with the
results obtained through the approach introduced in this work.

\section{The equations of motion}
\begin{figure}
	\begin{center}
		\includegraphics[width=0.44\textwidth]{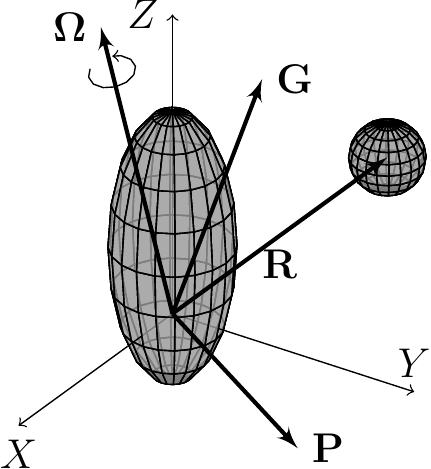}
	\end{center}
	\caption{Geometry of the problem. $\vX$, $\vY$, $\vZ$ are units vector coincided with the principal axis of inertia of the rigid body.}
\label{fig:fig1}
\end{figure}

Let us consider the gravitational two rigid body problem. We assume that one of
the bodies,  $\scP$ is a point mass, or an uniform sphere, with mass $m_{1}$.
The second one, $\scB$, is a rigid body with mass $m_{2}$
(Fig.~\ref{fig:fig1}).  The mechanical problem in the most general settings has
nine degrees of freedom, so the dynamics of the bodies are described by 18
first-order differential equations.  Moreover, the dynamical system possesses
symmetries, i.e., it may be shown that the equations of motion are invariant
with respect to \ed{a six-dimensional} group. This fact is a direct consequence
of the very basic laws of Newtonian mechanics. Namely, the equations of motion
do not depend on particular choice of the inertial reference frame. Thus, we can
choose the origin of the inertial frame in an arbitrary point, and this implies
that the equations of motion are invariant with respect to the natural action of
three dimensional group of translations. Moreover, also the orientation of the
inertial frame can be chosen arbitrarily.  Hence, the equations of motion are
invariant with respect to  the natural action of the three-dimensional group of
rotations.  

The existence of these symmetries \ed{makes it possible to} reduce the dimension
of the phase space of this system and to simplify its analysis.  However, there
is no unique procedure for such a reduction. Obviously, it should rely on such a
transformation of the initial system of the equations of motion (phase-space
variables) that the resulting equations form a lower-dimensional sub-system. In
fact, the reduced equations of motion of the two rigid body problem were
obtained by \cite{MR1130692} (actually, they considered  the motion of a rigid
body in the central gravitational field), and, with a simpler and more direct
method and under general settings by \cite{Maciejewski:00::c}. These papers are
good references to step-by-step development of the reduced equations of motion, 
which is skipped here to save space.

Before we write down these equations of motion, we need to fix the notation.
Components of a vector $\vx$ (as a geometric object) in an inertial reference
frame will be denoted by $\bx=[x_{1},x_{2},x_{3}]^{T}$. Components of the same
vector in the rigid body fixed frame we will 
denote by the corresponding capital letter, i.e.,
$\bX=[X_{1},X_{2},X_{3}]^{T}$. Thus, if $\bA$ is \ed{the orientation} (attitude) matrix of the
body with respect to the inertial frame, then we have $\bx=\bA\bX$.  The scalar
product of two vectors $\vx$ and $\vy$ is denoted by $\vx\cdot\vy$. It can be
calculated in an arbitrary orthonormal frame as follows:
\[
\vx\cdot\vy=\pairing{\bx}{\by}:=\sum_{i=1}^{3}x_{i}y_{i}=\pairing{\bA\bX}{\bA\bY}=\pairing{\bX}{\bY}=\sum_{i=1}^{3}X_{i}Y_{i}.
\]
We shall also write:
\[
x^{2}:=\vx\cdot\vx=\pairing{\bx}{\bx}=X^{2}:=\pairing{\bX}{\bX}.
\]
%
%% citation order
In our exposition, we follow \ed{\cite{Maciejewski:95::d,Maciejewski:00::c}}.   The
reduced equations of motion describe the relative motion of the bodies with
respect to a frame fixed in rigid body $\scB$. They have the following form
of the so-called {\ed{\em Newton-Euler equations}}: 
\begin{equation}
	\left.
	\begin{aligned}
		\Dt \bR &=\bR\times\bOmega + \bP,\\
		\Dt \bP&= \bP\times\bOmega - \pder{U}{\bR}, \\
		\Dt \bG&=\bG\times\bOmega +\bR\times  \pder{U}{\bR}, 
	\end{aligned}\quad\right\}
	\label{eq:em}
\end{equation} 
where $\bR$ is the radius vector directed from the mass centre of $\scB$ to
$\scP$,  $\bP$ is the relative linear momentum, $\bG$ is the angular momentum of
$\scB$, and $\bOmega=\bI^{-1}\bG$ is its angular velocity; $U$ is \ed{the
gravitational potential} of  
the body $\scB$,   and $\bI=\diag(A,B,C)$ stands for its tensor of inertia. We assume that
the rigid body fixed frame coincides with the principal axes frame. The geometry
of the model is illustrated in Fig.~\ref{fig:fig1}.

As it was shown in \cite{Maciejewski:00::c}, equations~\eqref{eq:em}
are Hamiltonian with respect to \ed{a non-canonical Poisson bracket}. They
admit two first integrals: the energy integral,
\begin{equation}
H=\frac{1}{2}\pairing{\bP}{\bP} +\frac{1}{2}\pairing{\bG}{\bOmega} + U(\bR),
\end{equation}
and the modulus of the total angular momentum $\bL$, which, \ed{with respect to} the body frame, 
is given by: 
\begin{equation}
L^{2}=\pairing{\bL}{\bL}, \mtext{where} \bL:=\bR\times\bP+\bG. 
\end{equation}
The total angular momentum with respect to the inertial  frame, i.e.,
\begin{equation}
\bl=\bA\bL,
\end{equation}
is a constant vector.  The time evolution of \ed{the attitude matrix} $\bA$ is given by the 
following kinematic equations: 
\begin{equation}
\Dt \bA = \bA\widehat\bOmega,
\end{equation}
where, as in  \cite{Maciejewski:00::c},  for a vector
$\bX=[X_{1},X_{2},X_{3}]^{T}$, we denote:
\begin{equation} 
\widehat\bX:=\begin{bmatrix}
\phantom{-}0 & -X_3 & \phantom{-}X_2\\ 
\phantom{-}X_3&  \phantom{-}0     & -X_1\\ 
-X_2& \phantom{-}X_1&\phantom{-}0 
\end{bmatrix}.
\end{equation}
If the rigid body $\scB$ does not possesses any additional symmetry, then the
relative equilibria of the considered two rigid body problem are just the
equilibria solutions (critical points) of the reduced equations~\eqref{eq:em}.
They were investigated by many authors, see \cite{MR1130692,
2006CeMDA..94..317S} and references therein.  The most intriguing result of
these investigations is the discovery of the  so-called {\em non-great-circle},
or {\em non-Lagrangian} solutions. If the system is in a relative  equilibrium,
then the relative orbit is circular. However, in a case of  non-Lagrangian
equilibrium, the orbital planes of the point $\scP$  and  the rigid body $\scB$
are mutually parallel but do not coincide.  These classes of equilibria have
been discovered by \cite{1979AZh....56..881A}.  Later on, they were
re-discovered by \cite{MR1130692}.

\section{The symmetric Kinoshita problem}
From hereafter, we shall assume that the rigid body $\scB$ is dynamically
symmetric with respect to its third principal axis, i.e., we set
$\bI=\diag(A,A,C)$. Moreover, we assume also that the gravity field of $\scB$ is
symmetric with respect to the same axis and that it can be expressed 
through the following condition:
\begin{equation}
  \label{eq:symcon}
  R_{1}\frac{\partial\, U}{\partial R_{2}}(\bR) -  
 R_{2}\frac{\partial\, U}{\partial R_{1}}(\bR)=0.
\end{equation}
(We note, for a further reference, that {\em the equatorial plane}
of the rigid body is defined in such a way, that it contains the mass centre 
and this body and the symmetry axis is normal to that plane).
Under these assumptions, equations~\eqref{eq:em} have an additional first
integral $G_{3}$. The existence of this first integral is connected with the
fact that now the reduced equations~\eqref{eq:em} have one-dimensional symmetry
group.   To show this explicitly, we need to introduce additional notations and
to state or recall some elementary facts. 

By $\mathrm{SO}(3,\R)$ we denote the special orthogonal group, i.e.,
the group of $3\times 3$ matrices $\bA$, satisfying $\bA\bA^{T}=\bE$, and
$\det \bA=1$. Its Lie algebra $\mathrm{so}(3,\R)$ consists of $3\times
3$ antisymmetric matrices. It can be identified with $\bR^{3}$ by the
following isomorphism:
\begin{equation}
  \nonumber
  \R^{3}\ni \bX\longmapsto \widehat\bX:=\begin{bmatrix}
\phantom{-}0 & -X_3 & \phantom{-}X_2\\ 
\phantom{-}X_3&  \phantom{-}0     & -X_1\\ 
-X_2& \phantom{-}X_1&\phantom{-}0 
\end{bmatrix}\in \mathrm{so}(3,\R). 
\end{equation}
Later on, we use several identities which we collect as the following
\ed{statements \citep[e.g.,][]{Maciejewski:95::d}}.
\begin{proposition}
\label{pro:hat}
For arbitrary $\bX$, $\bY\in\R^{3}$, and $\bA\in\mathrm{SO}(3,\R)$, the
following identities are hold true:
\begin{enumerate}
	\item $ {\widehat\bX}^T = -{\widehat\bX}={\widehat{(-\bX)}}$.
	\item \label{m:cross} $ {\widehat\bX}\bX = \bX\times\bY = -{\widehat\bY}\bX$. 
	\item ${\widehat\bX}{\widehat\bY}=\bY\bX^T - \bX^T\bY \bE$. 
	\item $\widehat{\bX\times\bY}= {\widehat\bX}{\widehat\bY}-{\widehat\bY}{\widehat\bX}$.
	\item $\widehat{\bA\bX}=\bA\widehat\bX\bA^T$.
	\item $\bA(\bX\times\bY)=\bA\widehat{\bX}\bY=(\bA\bX)\times(\bA\bY)$,
\end{enumerate}
where $\bE$ denotes the $3\times 3$ identity matrix. 
\end{proposition}
All rotation about the third axis of the body form a subgroup of
$\mathrm{SO}(3,\R)$ isomorphic with $\mathrm{SO}(2,\R)$.  We show that
equations~\eqref{eq:em} possess $\mathrm{SO}(2,\R)$ symmetry.  That is,  they do
not change its  form after the following transformation:
\begin{equation}
\label{eq:ch}
[\bR,\bP,\bG]=[\bA_{3}\bR^{*}, \bA_{3}\bP^{*},\bA_{3}\bG^{*}],
\end{equation}
where $\bA_{3}$ is an arbitrary  matrix of rotation about the third
axis of the body, i.e., a matrix of the following form:
\begin{equation}
\label{eq:r3}
\bA_{3}=\begin{bmatrix}
a & -b & 0\\
b & \phantom{-}a& 0\\
0 & \phantom{-}0& 1
\end{bmatrix}\mtext{where} a^{2}+b^{2}=1. 
\end{equation}
\ed{Let us notice} that $\bA_{3}\bI=\bI\bA_{3}$ \ed{(i.e.,$\bI$ commutes with
$\bA_{3}$), hence we have},
\[ 
\Dt \bR =\bA_{3}\Dt \bR^{*} =(\bA_{3}\bR^{*})\times
(\bI^{-1}\bA_{3}\bG^{*})+
\bA_{3}\bP^{*}.
\]
\ed{This equation may be rewritten as}
\[
(\bA_{3}\bR^{*})\times
(\bI^{-1}\bA_{3}\bG^{*})=(\bA_{3}\bR^{*})\times
(\bA_{3}\bI^{-1}\bG^{*})=\bA_{3}\left[ \bR^{*}\times
\bI^{-1}\bG^{*}\right],
\]
\ed{where the last equality} follows from the sixth identity given in
Proposition~\ref{pro:hat}. Hence,
\[ 
\bA_{3}\Dt \bR^{*} =\bA_{3}\left[\bR^{*}\times
\bOmega^{*}+
\bP^{*}\right], 
\]
where $\bOmega^{*}=\bI^{-1}\bG^{*}$, \ed{and, finally,}
\[ 
\Dt \bR^{*} =\bR^{*}\times
\bOmega^{*}+
\bP^{*}, 
\]
and this shows that the first equation in~\eqref{eq:em} is invariant with
respect to variables change~\eqref{eq:ch}. In a similar manner, we show that the
remaining two equations have the same property. 
\section{Relative equilibria}

\subsection{General theory of the relative equilibria}
Let us recall a formal, geometric definition of a relative equilibrium of a
system with symmetry \cite[see, for instance][]{Marsden:94::},
which is crucial to perform our analysis: {\em a relative equilibrium is a
solution of the system represented by the phase-space curve which is an orbit of
a point under the action of a one-dimensional subgroup of the symmetry group of
the system}.

In our case, the symmetry group is $\mathrm{SO}(2,\R)$ which is identified with
matrices of the form~\eqref{eq:r3}. It is a one-dimensional group. Let us put
$a=\cos\varphi$ and  $b=\sin\varphi$ in equation~\eqref{eq:r3}. With this parameterization, we denote elements of $\mathrm{SO}(2,\R)$ by $\bA_{3}(\varphi)$.   Hence, an orbit of a
point $[\bR_{0}, \bP_{0}, \bG_{0}]^{T}$ under the action of $\mathrm{SO}(2,\R)$ 
is the following set:
\begin{equation}
  \nonumber
\defset{[\bA_3(\varphi)\bR_0,\bA_3(\varphi)\bP_0,\bA_3(\varphi)\bG_0]\in\R^{9}}{\varphi\in[0,2\pi)},
\end{equation}
\ed{where $\bR_{0}, \bP_{0}, \bG_{0}$ are constant vectors.}
Thus, the relative equilibrium of the symmetric Kinoshita problem is a solution
to equations~\eqref{eq:em}, which has the following general form:  
\begin{equation} 
	\left.
	\begin{aligned}
		\bR(t) &= \bA_3(Nt)\bR_0,\\
		\bP(t) &= \bA_3(Nt)\bP_0,\\
		\bG(t) &= \bA_3(Nt)\bG_0,\\
	\end{aligned}
	\quad\right\}
	\label{eq:sol}
\end{equation}
where $N$ is \ed{the real number}. Notice, that
$\bOmega(t)=\bI^{-1}\bG(t)=\bA_{3}(Nt)\bOmega_{0}$, where
$\bOmega_{0}=\bI^{-1}\bG_{0}$. In other words, the relative equilibrium is a
special periodic solution to the equations of motion \eqref{eq:em}.   From the
above formulae we can deduce more useful conclusions on the relative
equilibrium:
	\begin{enumerate}
		\item vectors $\bR(t)$, $\bP(t)$, $\bG(t)$ and $\bOmega(t)$ have
                  constant lengths,
		\item their ``third'' components $R_{3}(t)$, $P_{3}(t)$,
                  $G_{3}(t)$ and $\Omega_{3}(t)$ are constant,
		\item all angles between vectors $\bR(t)$, $\bP(t)$,
                  $\bG(t)$ and $\bOmega(t)$ are constant.
	\end{enumerate}
Let us notice that the relative equilibrium \eqref{eq:sol} does not determine 
an unique set of $[\bR_{0},\bP_{0},\bG_{0}]$. In fact, instead of
$[\bR_{0},\bP_{0},\bG_{0}]$, we can take
$[\bA({\varphi})\bR_{0},\bA({\varphi})\bP_{0},\bA({\varphi})\bG_{0}]$, with
arbitrary chosen constant $\varphi$.  One would like to perform a reduction of
\eqref{eq:em} in such a way that in the reduced system, the relative equilibrium
corresponds to \ed{an ``usual''} equilibrium point (i.e., \ed{which is} understood as the
critical point of the equations of motion).  It is possible, however, it may
lead to singularities. For example, let us introduce
\ed{three sets of} cylindrical coordinates:
\begin{equation}
	\left.
	\begin{aligned}
		\bR &= \left( \rho\cos\varphi,\rho\sin\varphi, R_{3} \right)^T,\\
		\bP &= \left( p\cos\upsilon,p\sin\upsilon,P_3 \right)^T,\\
		\bOmega &= \left( \omega\cos\psi,\omega\sin\psi,\Omega_3 \right)^T.
	\end{aligned}
	\quad\right.
	\nonumber
\end{equation}
Using these coordinates, we may write:
\begin{equation}\nonumber
 \pder{U}{\bR} = \left( U_\rho\cos\varphi,U_\rho\cos\varphi,U_3 \right)^T,
\end{equation}
where, 
\[
U_\rho=\pder{U}{\rho}, \mtext{and} U_{3}=\pder{U}{R_{3}}.
\]
With respect to these new variables, system~\eqref{eq:em} reads as follows:
\begin{equation}
	\left.
	\begin{aligned}
		\Dt\rho &= p\cos(\upsilon-\varphi)-z\omega\sin(\psi-\varphi),\\
		\Dt z &= P_3+\rho\omega\sin(\psi-\varphi),\\
		\Dt\upsilon &= p^{-1}\left[ P_3\omega\cos(\psi-\varphi)+U_\rho\sin(\psi-\varphi)-p\Omega_3 \right],\\
		\Dt p &= -P_3\omega\sin(\psi-\varphi)-U_\rho\cos(\upsilon-\varphi),\\
		\Dt P_3 &= p\omega\sin(\psi-\upsilon)-U_3,\\
		\Dt\omega &= A^{-1}\left[ zU_\rho-\rho U_3 \right]\sin(\psi-\varphi),\\
		\Dt\varphi &= \rho^{-1}\left[ p\sin(\upsilon-\varphi)+z\omega\cos(\psi-\varphi)-\rho\Omega_3 \right],\\
		\Dt\psi &= (A\omega)^{-1}\left[ \left( zU_\rho-\rho U_3 \right)\cos(\psi-\varphi)-(A-C)\omega\Omega_3 \right],
	\end{aligned}
	\quad\right\}
	\nonumber
\end{equation}
Hence, finally we have one equation less than in~\eqref{eq:em} because
$\Omega_3$ is constant.  Note that the right hand  sides of the equations depend
on the difference between the angles but not on the angles themselves. Thus we
can introduce the following new set of variables:
\begin{equation}
		\alpha = \psi-\varphi,\qquad 
		\beta = \upsilon-\varphi,
	\nonumber
\end{equation}
and then we obtain: 
\begin{equation}
		\left.
	\begin{aligned}
		\Dt\rho &= p\cos(\beta)-z\omega\sin(\alpha),\\
		\Dt z &= P_3+\rho\omega\sin(\alpha),\\
		\Dt p &= -P_3\omega\sin(\alpha-\beta)-U_\rho\cos(\beta),\\
		\Dt P_3 &= p\omega\sin(\alpha-\beta)-U_3,\\
		\Dt\omega &= \left[ zU_\rho-\rho U_3 \right]\sin(\alpha),\\
		\Dt\alpha &= \left( zU_\rho-\rho U_3
                \right)\omega^{-1}\cos(\alpha)+\left( C\Omega_3 -\right.\\
		& \left. A\omega \right)\left[ p\sin(\beta)+z\omega\cos(\alpha)\right]\rho^{-1},\\
		\Dt\beta &= p^{-1}\left[ P_3\omega\cos(\alpha-\beta)+U_\rho\sin(\beta) \right]-\\
		& \rho^{-1}\left[ p\sin(\beta)+z\omega\cos(\alpha) \right],\\
		 \Dt\varphi &= 0.\\
	\end{aligned}
	\quad\right\}
	\nonumber
\end{equation}
The first seven equations in the above set represent a system of equations
which  flows from  the reduction~\eqref{eq:em} by the symmetry. Now, a relative
equilibrium of \eqref{eq:em} corresponds to an equilibrium point of the above
system.  However, this system has singularities when one of variables $\rho$,
$\omega$ or $p$ vanishes.  Because of this fact it would be very difficult to
perform global, qualitative analysis which is our primary goal. 

\subsection{The geometric reduction of the system}
A possibility that the reduction may introduce singularities to the equations of
motion inspired us  and, in fact, forced us to choose \ed{a particular} approach which
we describe below in detail. Clearly, instead of using any local variables, it
is much more convenient to derive the necessary and sufficient conditions for
the existence of  relative equilibria in terms of the original global and
non-singular variables $(\bR,\bP,\bG)$. 

Let us assume that \ed{a relative} equilibrium is given by~\eqref{eq:sol}. Our aim is
to find equations determining $(\bR_{0},\bP_{0},\bG_{0})$, and $N$. At first, we
may notice that: 
\begin{equation}
	\Dt \bA_3(Nt)=\bA_3(Nt)\widehat{\bN},
	\nonumber
\end{equation}
where $\bN=\left[0,0,N \right]^T$. Thus we \ed{may easily} derive that:
\begin{equation}
  \nonumber
  \Dt \bR(t)=\Dt \bA_3(Nt) \bR_{0}=\bA_3(Nt)\widehat{\bN} \bR_{0}=\bA_3(Nt)(\bN\times\bR_{0}).
\end{equation}
On the other hand, we have also: 
\[
  \Dt \bR(t)= \bR(t)\times \bOmega(t) +\bP(t)= \bA_3(Nt)(
  \bR_{0}\times \bOmega_{0}+\bP_{0}),
\]
as the system has $\mathrm{SO}(2,\R)$ symmetry. So, we have
\begin{equation}
	\bN\times\bR_0 = \bR_0\times\bOmega_0+\bP_0.
	\nonumber
\end{equation}

We proceed in a similar way with other components of the phase variables
defining the relative equilibrium,
i.e., with $\bP(t)$ and $\bG(t)$.  Finally,  we obtain the following
equations:
\begin{equation}
\nonumber
	\left.
	\begin{aligned}
		\bN\times\bR_0 &= \bR_0\times\bOmega_0+\bP_0,\\
		\bN\times\bP_0 &= \bP_0\times\bOmega_0-\pder{U}{\bR_0},\\
		\bN\times\bG_0 &= \bG_0\times\bOmega_0+\bR_0\times
		\pder{U}{\bR_0}.\\
	\end{aligned}
	\quad\right\}
	\nonumber
\end{equation}
This system  can be rewritten in a more compact form as follows:
\begin{equation}
	\left.
	\begin{aligned}
		\widetilde{\bOmega}\times\bR_0 &= \bP_0,\\
		\widetilde{\bOmega}\times\bP_0 &= -\pder{U}{\bR_0},\\
		\widetilde{\bOmega}\times\bG_0 &= \bR_0\times \pder{U}{\bR_0},\\
	\end{aligned}
	\quad\right\}
\label{eq:cr}
\end{equation}
where,
$
	\widetilde{\bOmega} = \bOmega_0 + \bN,
$
is \ed{the instantaneous} angular velocity of the rigid body.

System~\eqref{eq:cr} defines {\em nine equations} which must be satisfied by
{\em ten unknowns}.  We found that there are several ways to overcome this
inconvenience. In our further procedure, we choose the following 
approach.  We look
for the relative equilibria which exist for a given value of the angular
momentum of the system, chosen as the model parameter.  Because 
\begin{equation}
	L^2=\left( \bR\times\bP+\bG \right)^2 = \left( \bR_0\times\bP_0+\bG_0 \right)^2,
	\nonumber
\end{equation}
after the elimination of $\bP_0$, we obtain:
\begin{equation}
	L^2 = \left( \widetilde{\bOmega}R_0^2-\bR_0\langle
          \widetilde{\bOmega},\bR_0 \rangle+\bG_0 \right)^2.
	\nonumber
\end{equation}
Hence, our goal is to find all solutions \ed{to} the following nonlinear
equations: 
\begin{subequations}
	\begin{align}
\label{e1}
          \widetilde{\bOmega}\times\bR_0 &= \bP_0,\\
\label{e2}
          \widetilde{\bOmega}\langle \widetilde{\bOmega}, \bR_0
          \rangle- \bR_0 \widetilde{\Omega}^2 &= -\pder{U}{\bR_0},\\
\label{e3}
          \widetilde{\bOmega}\times\bG_0 &= \bR_0\times
          \pder{U}{\bR_0},\\
\label{e4}
          \left( \widetilde{\bOmega}R_0^2-\bR_0\langle \widetilde{\bOmega},\bR_0 \rangle + \bG_0 \right)^2 &= L^2.
	\end{align}
	\label{eq:sys_fullnoP}
\end{subequations}
In the above equations,  $L$, and the third component of $\bG_{0}$ are now fixed
parameters of the reduced model. 
\subsection{Geometric description of the equilibria solutions}
The equations of motion~\eqref{eq:em} describe the relative motion of the bodies in
the rigid body fixed frame. Apparently,  this \ed{makes it difficult to
obtain the} geometrical and physical interpretation. In fact, we can deduce all necessary information quite
easily. 

First of all, let us recall that the total angular momentum $\vl$ is a constant
vector. So, we can choose an inertial frame in such a way that its third axis is
directed along that vector. If $\bR(t)=\bA_{3}(Nt)\bR_{0}$ describes the
relative orbit in a relative equilibrium, and $\vr(t)$ is the corresponding
equilibrium vector, then we have: 
	\begin{equation*}
	\vl\cdot\vr(t)=\pairing{\bl}{\br(t)}=\pairing{\bA\bL}{\bA\bA_{3}(Nt)\bR_{0}}=
	\pairing{\bL}{\bA_{3}(Nt)\bR_{0}}.
	\end{equation*}
Moreover, the following general relation holds true:
\[
\bL=\bA_{3}(Nt)( \bR_{0}\times \bP_{0}+\bG_{0}),
\]
so, finally, we have:
	\begin{equation}
	\vl\cdot\vr(t)=\pairing{ \bR_{0}\times \bP_{0}+\bG_{0}}{\bR_{0}}
	=\pairing{\bG_{0}}{\bR_{0}}.
		\label{eq:orbit_test}
	\end{equation}
We can conclude that in a relative equilibrium, the projection of the relative
radius vector onto the total angular momentum vector is constant. An important
conclusion follows immediately: if this projection is non-zero, then {\em orbits of
the point and the rigid body lie in different planes}. 

Let $\vz(t)$ denotes the unit vector along the symmetry axis of the body. Then 
\begin{equation}
  \nonumber
  \vl\cdot \vz(t)=\pairing{\bL}{\bZ}=
\pairing{\bA_{3}(Nt)( \bR_{0}\times \bP_{0}+\bG_{0})}{\bZ}=
\pairing{\bR_{0}\times \bP_{0}+\bG_{0}}{\bZ},
\end{equation}
because $\bZ=[0,0,1]^{T}$, and so, $\bA_{3}(Nt)\bZ=\bZ$. Thus, in a
relative equilibrium, the projection of the symmetry axis onto the
total angular momentum is also constant.

Let us introduce the orbital reference frame with axes parallel to $\vr(t)$,
$\dot\vr(t)$ and $\vc(t):=\vr(t)\times \dot\vr(t)$. \ed{Because} the relative orbit in a
relative equilibrium  is circular, this frame is orthogonal.  We show that the
projection of the symmetry axis $\vz(t)$ onto the axis of this frame is also
constant. In other words, we are going to prove that in the relative equilibrium
the symmetry axis of the rigid body has a fixed orientation with respect to the
orbital reference frame. 

To shorten notation, we shall write $\bA_{3}$ instead \ed{of} $\bA_{3}(Nt)$. Let
$\bB=\bA\bA_{3}$. Then we have:
\[
\Dt\bB= \dot \bA \bA_{3} +\bA\dot\bA_{3}=
\bA\widehat\bOmega\bA_{3}+\bA\bA_{3}\widehat\bN.
\]
But in a relative equilibrium we have $\bOmega=\bA_{3}\bOmega_{0}$. By the fifth
property given in Proposition~\ref{pro:hat}:
\[
\widehat\bOmega=\bA_{3}\bOmega_{0}\bA_{3}^{T},
\]
hence we obtain that:
\[
\Dt\bB=
\bB\widehat{\widetilde\bOmega},
\] 
where $\widetilde\bOmega=\bOmega_{0}+\bN$ is a constant vector. Now,
it is easy to show that:
\begin{equation}
\begin{aligned}
  \br(t) &= \bA\bR(t)=\bB\bR_0,\\
  \dot{\br}(t) &= \bB( \widetilde\bOmega\times\bR_{0}),\\
  \br(t)\times\dot{\br}(t) &= \bB( \bR_{0}\times(
  \widetilde\bOmega\times\bR_{0})). 
		\end{aligned}
\end{equation}
Because $\bz(t)=\bA\bZ=\bB\bZ$, and $\bA_{3}\bZ=\bZ$, we have:	
\begin{equation}
		\begin{aligned}
	\vr(t)\cdot\vz(t)
        &=\pairing{\br(t)}{\bz(t)}=\pairing{\bB\bR_{0}}{\bB\bZ}=\pairing{\bR_{0}}{ \bZ},\\
	\dot\vr(t)\cdot\vz(t)& = \pairing{
          \widetilde\bOmega\times\bR_{0}}{\bZ}=\pairing{\bP_{0}}{\bZ}, \\
\vc(t)\cdot\vz(t)&= \pairing{ \bR_{0}\times(
  \widetilde\bOmega\times\bR_{0})}{\bZ}.
		\end{aligned}
		\label{orb}
	\end{equation}
The above formulae prove our initial claim.

\section{Conditions for the relative equilibria}
\label{s:solutions}
In this section, we show that all relative equilibria can be determined by
solutions of certain non-linear system comprising of only two non-linear 
{\em scalar} equations. We shall also prove that there exist {\em three classes} of these
relative equilibria in the dynamical model which we consider.

In order to simplify notation while working with coordinates,  we  shall
\ed{skip index ``0'' when denoting coordinates of vectors $\bR_{0}$, $\bP_{0}$,
$\bG_{0}$ from hereafter}. Thus,  we set 
\begin{equation}
	\left.
	\begin{aligned}
		\bR_0 &= \left[ R_1, R_2, R_3 \right]^T,\\
		\bP_0 &= \left[ P_1, P_2, P_3 \right]^T,\\
		\bOmega_{0} &= \left[\Omega_1, \Omega_2, \Omega_3 \right]^T,\\
		\widetilde{\bOmega} &= \left[ \Omega_1, \Omega_2, \Omega_3+N \right]^T,\\
		\bG_0 &= \left[ A\Omega_1, A\Omega_2, C\Omega_3 \right]^T,\\
		\pder{U}{\bR_{0}}&:=\pder{U}{\bR}(\bR_{0}) = \left[ U_1, U_2, U_3 \right]^T.\\
	\end{aligned}
	\quad\right.
	\nonumber
\end{equation}
The fact that the body is symmetric implies that: 
\begin{equation}
   R_1U_2-R_2U_1=0,
	\label{eq:ax_pot}
\end{equation}
with accord to~\eqref{eq:symcon}.  Taking the vector product of both sides of
equation~\eqref{e2} with $\bR_{0}$, we obtain: 
\begin{equation}
  \label{eq:r}
  \pairing{\widetilde\bOmega}{\bR_{0}}\widetilde\bOmega\times\bR_{0}=\bR_{0}\times	\pder{U}{\bR_{0}}.
\end{equation}
Hence, taking into account~\eqref{eq:ax_pot}, we have:
\begin{equation}
  \nonumber
  \pairing{\widetilde\bOmega}{\bR_{0}}
\pairing{\bZ}{\widetilde\bOmega\times\bR_{0}}=0.
\end{equation}
Thus, we have to consider two different cases. 
  
If $\pairing{\widetilde\bOmega}{\bR_{0}}=0$  then from~\eqref{eq:r} it
follows immediately that:
\[
\bR_{0}\times	\pder{U}{\bR_{0}}=\bzero,
\]
or, $\pairing{\bZ}{\widetilde\bOmega\times\bR_{0}}=0$, and this gives
\[
\Omega_1R_2-\Omega_2R_1=0.
\]
In the first case, the gradient of $U$ at point $\bR_{0}$ is parallel to the
radius vector. According \ed{to} \cite{2006CeMDA..94..317S}, such a point is called
{\em locally central}. We shall say that a relative equilibrium is locally
central if point $\bR_{0}$ is locally central.
 
In our analysis, we assume that the gradient of the potential does not vanish at
a considered point $\bR_{0}$. Thus,  a relative equilibrium is 
{\em locally central}
if and only if  $\pairing{\widetilde\bOmega}{\bR_{0}}=0$.  In fact, if $\bR_{0}$
is  locally central, then from~\eqref{eq:r} it follows that either 
$\pairing{\widetilde\bOmega}{\bR_{0}}=0$, or
$\widetilde\bOmega\times\bR_{0}=\bzero$. If the second possibility
occurs, then, by~\eqref{e1}, $\bP_{0}=\bzero$, and, as equations~\eqref{eq:cr}
imply, the gradient of $U$ at $\bR_{0}$ vanishes.  But, according with our
assumptions, it is impossible. 

At the final stage of our analysis, it is convenient to use cylindrical
coordinates $(\rho, \varphi,R_{3})$ instead of $(R_{1},R_{2},R_{3})$.  The
potential $U$ expressed with respect to the cylindrical coordinates depends only
on $\rho$, and $R_{3}$; still, as a function of these two arguments it will be
denoted by the same symbol $U$.  It will be clear from the context which
coordinates are in use.

\subsection{Locally central relative equilibria}
\label{s:lc}
If $\bR_{0}$ is locally central, then, as we have shown,  $\langle
\widetilde{\bOmega},\bR_0 \rangle=0$, and of course:
\[
\bR_{0}\times	\pder{U}{\bR_{0}}=\bzero.
\]
Thanks to this fact, the basic system of the equations of motion
\eqref{eq:sys_fullnoP} may be simplified considerably, and it  reads as
follows:
\begin{subequations}
	\begin{align}
\label{c1}
         \bP_{0}&= \widetilde{\bOmega}\times\bR_0 ,\\
\label{c2}
         \bR_0 \widetilde{\Omega}^2 &= \pder{U}{\bR_0},\\
\label{c3}
          \widetilde{\bOmega}\times\bG_0 &= \bzero,\\
\label{c4}
          \left( \widetilde{\bOmega}R_0^2 + \bG_0 \right)^2 &= L^2.
	\end{align}
	\nonumber
\end{subequations}
Equation~\eqref{c3} is equivalent to the following two scalar equations:
\begin{equation}
	\left.
	\begin{aligned}
		\left[ A\widetilde{\Omega}_3-C\Omega_3 \right]\Omega_2 &= 0,\\
		\left[ A\widetilde{\Omega}_3-C\Omega_3 \right]\Omega_1 &= 0.\\
	\end{aligned}
	\quad\right\}
	\nonumber
\end{equation}
Therefore, either  $\Omega_1=\Omega_2=0$, or
$A\widetilde{\Omega}_3-C\Omega_3=0$. We investigate both cases separately.

\subsubsection{Case $\Omega_1=\Omega_2=0$ \ed{(cylindrical precession)}}
\label{s:cylindrical}
If $\Omega_1=\Omega_2=0$ then  condition
$\pairing{\widetilde\bOmega}{\bR_{0}}=0$ reduces to
$\widetilde\Omega_{3}R_{3}=0$. But $\widetilde\Omega_{3}\neq 0$,  otherwise
$\widetilde{\Omega}^2=0$, and, as~\eqref{c2} shows, the gradient of $U$
vanishes and then $R_{3}=0$. Moreover, $R_{1}$ and $R_{2}$ do not vanish
simultaneously, otherwise $\bR_{0}=\bzero$. Hence, we can safely  use
cylindrical coordinates \ed{$(\rho,z\equiv 0)$\footnote{Note that due to the
symmetry, the third angular coordinate $\phi$ is irrelevant here.}} 
because $\rho> 0$. We introduce two functions of $\rho$ given by:
\begin{equation}
  \nonumber
  f(\rho):=U_{3}(\rho,0),  \mtext{and} g(\rho):=U_{\rho}(\rho,0). 
\end{equation}
Now, equations \eqref{c2} and \eqref{c4} lead to the following system:
\begin{equation}
  f(\rho)=0, \qquad 
  \rho\widetilde{\Omega}_3^2 =
  g(\rho),\qquad
  \left[ \rho^2
    \widetilde{\Omega}_3+C\Omega_3
  \right]^2 = L^2
\end{equation}
Through appropriate simplification, we find that the above system is equivalent to
the following ones:
\begin{equation}
  f(\rho)=0, \qquad \left( L+\varepsilon C\Omega_3 \right)^2 = \rho^3
  g(\rho), \qquad  \widetilde{\Omega}_3=	-\varepsilon
  \frac{L+\varepsilon C\Omega_3}{\rho^2},
\end{equation}
where parameter $\varepsilon\in\{-1,1\}$. 

\begin{figure}
	\begin{center}
		\includegraphics[width=0.44\textwidth]{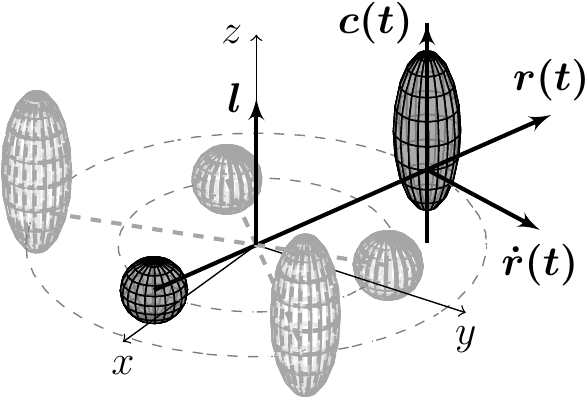}
	\end{center}
	\caption{\textit{Cylindrical precession} in the inertial reference frame. Shaded bodies show the attitude of the rigid body $\scB$ and the point $\scP$ in other 
\ed{moments of time}.}
	\label{fig:fig2}
\end{figure}

The above considerations can be summarized in the following way. In the
considered case, the relative equilibrium exists if and only if there
exists a solution $\rho>0$ of equations:
\begin{equation}
\label{e-cyl}
  f(\rho)=0, \qquad \left( L+\varepsilon C\Omega_3 \right)^2 = \rho^3
  g(\rho). 
\end{equation}
If such a solution exists then we define: 
\[
R_{1}=\rho\cos \varphi, \quad R_{2}=\rho\sin \varphi, \qquad
\alpha^{2}=\rho^{3}U_{\rho},\quad \beta=\frac{\alpha}{\rho^{2}},
\]
where  $\varphi\in [0, 2\pi]$ can be chosen arbitrarily.  
Then the relative equilibrium is determined through:
\begin{equation}
\bR_{0}=  \begin{bmatrix}R_{1}\\R_{2}\\ 0
\end{bmatrix}, \quad
\bP_{0}=\beta \begin{bmatrix} -R_{2}\\\phantom{-} R_{1}\\ 0
\end{bmatrix},
\quad 
\bG_{0}=C\begin{bmatrix}0\\ 0\\ \Omega_{3}
\end{bmatrix},
\quad N=\beta-\Omega_{3}.\nonumber
\end{equation}
From the above formulae it immediately follows that  point $\scP$ and the mass
centre of the rigid body $\scB$ move in one plane. In fact,
by~\eqref{eq:orbit_test}, we have: 
\begin{equation}
  \vl\cdot\vr(t)
  =\pairing{\bG_{0}}{\bR_{0}}=0.
  \nonumber
\end{equation}
Moreover, by \eqref{orb}, we have also: 
\begin{equation}
		\begin{aligned}
	\vr(t)\cdot\vz(t)
        &=\pairing{\bR_{0}}{ \bZ}=0,\\
	\dot\vr(t)\cdot\vz(t)& =\pairing{\bP_{0}}{\bZ}=0.
		\end{aligned}
		\nonumber
	\end{equation}
Hence, the axis of symmetry of the rigid body is perpendicular to the plane of the
relative orbit, see Fig.~\ref{fig:fig2}. We called this kind of the relative equilibrium 
as {\em the cylindrical precession} from hereafter. In the paper of Kinoshita,
it is named, after \cite{1959SvA.....3..154D}, as {\em the float solution}. 
Regarding existence conditions of this solution, let us notice that
system~\eqref{e-cyl} is overdetermined, we have two equations for one variable
$\rho$. However, if the body is symmetric with respect to the equatorial
plane, then $U$
is an even function of $R_{3}$. In this instance, we have:
\begin{equation}
  \nonumber
  f(\rho)=\pder{U}{\bR}(\rho,0) \equiv 0,
\end{equation}
identically. Thus, in this case, the  cylindrical precession exists  if and only
if the second equation of system~\eqref{e-cyl} has a solution. Moreover, if the
potential is such that there exists $\rho>0$, for which $g(\rho)>0$, then we
always may find parameters of the problem such that equations~\eqref{e-cyl} are
fulfilled. 

\begin{figure}
	\begin{center}
		\includegraphics[width=0.44\textwidth]{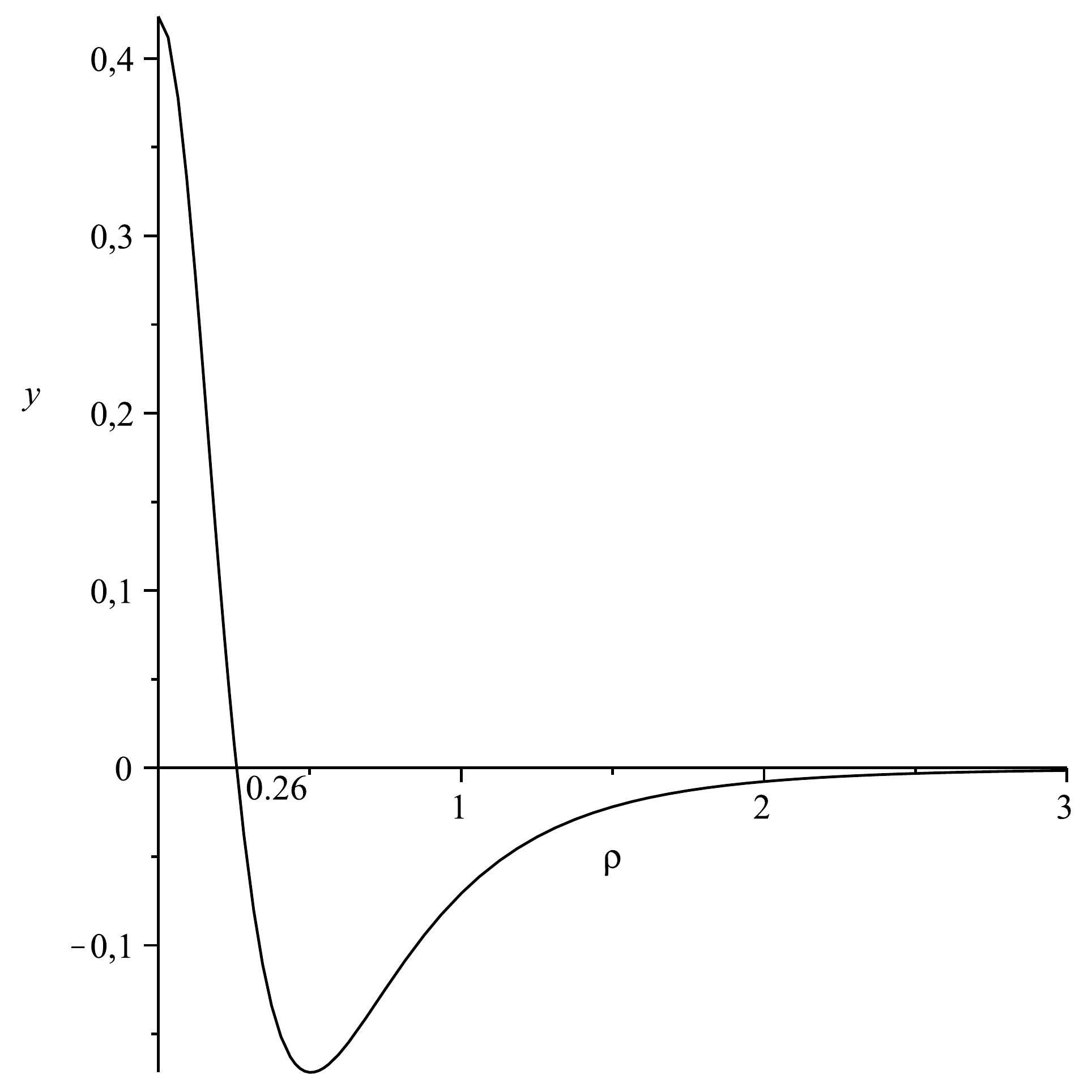}
	\end{center}
	\caption{
The plot of $f(\rho)$ crosses $\rho$-axis for $\rho\simeq 0.26$. The values
of parameters are chosen in the following way: $\mu_1=0.15$, $\mu_2=0.3$,
$\mu_3=0.55$, where $\mu_i={m_1}/{\sum_{j=1}^{3}m_j}$ and $m_i$ is mass
of the $i$-th sphere. See the text for details.}
	\label{fig:fig3}
\end{figure}

If the body is not symmetric with respect to the equatorial plane, the first equation of
system ~\eqref{e-cyl} is not satisfied identically.  Thus a question  emerges:
does there exist an axially  symmetric body which is {\em not} symmetric with
respect to the equatorial plane, and for which equation:
\[
U_{3}(\rho,0)=0, 
\]
has a solution $\rho>0$?  To answer this question, let us consider an
axisymmetric rigid body composed of three uniform spheres whose centers of
masses are placed at the same line.  The masses of these spheres are chosen
in \ed{a way} to guarantee that the potential is not symmetric with respect
to the equatorial plane.  The resulting function $f(\rho)$ is depicted in
Fig.~\ref{fig:fig3}.  Since \ed{its} graph crosses the $\rho$-axis for $\rho>0$,
there \ed{exist such points which satisfy the first condition of system}
~\eqref{e-cyl}, indeed.  Then the parameters of the system can
always be chosen \ed{to fulfill} the last condition of~\eqref{e-cyl}.  Thus, the
cylindrical precessions may exist in the system with the rigid body, {\em
which is not symmetric with respect to its equatorial plane}.

\subsubsection{Case $A\widetilde{\Omega}_3=C\Omega_3$ \ed{(inclined co-planar precession)}}
\label{s:inclined_planar}
Condition $A\widetilde{\Omega}_3=C\Omega_3$ implies immediately that 
\begin{equation}
  \nonumber
  N=\frac{C-A}{A}\Omega_{3}.
\end{equation}
Moreover, now $\pairing{\widetilde{\bOmega}}{\bR_{0}}=0$, is equivalent to
$\pairing{{\bG}_{0}}{\bR_{0}}=0$, because $A\widetilde{\bOmega}=\bG_{0}$.
We have
even more, because now equation~\eqref{c2} can be rewritten in the following
form:
  \begin{equation}
    \label{eq:p0}
    \bP_{0}=\frac{1}{A}\bG_{0}\times\bR_{0}.
  \end{equation}
and equation~\eqref{c4} reads as:
\begin{equation}
  \label{eq:c4e}
  L^{2}=  \left( \widetilde{\bOmega}R_0^2 + \bG_0 \right)^2 =\frac{G_{0}^{2}}{A^{2}}(R_{0}^{2}+A)^{2}.
\end{equation}
Therefore, equation~\eqref{c2} can be rewritten as follows:
\begin{equation}
\label{ce2}
  \frac{L^2}{(R_{0}^{2}+A)^{2}}\bR_{0}=\pder{U}{\bR_{0}}.
\end{equation}
A solution of this equation gives $\bR_{0}$. However, this form is not
convenient for further analysis because solutions to the above equation, 
\ed{whenever they exist}, are not isolated. In terms of \ed{the cylindrical} coordinates,
we may rewrite~\eqref{ce2} as follows:
\begin{equation}
  \left.
    \begin{aligned}
      \rho\frac{L^2}{\left( R_3^2+\rho^2+A \right)^2} &= U_\rho,\\
      R_3\frac{L^2}{\left( R_3^2+\rho^2+A \right)^2} &= U_3,\\
    \end{aligned}
    \quad\right\}
  \label{eq:spoke_get_key}
\end{equation}
A solution to the above system gives us $\bR_{0}$. Because,
\ed{due to the symmetry}, we may choose freely
the polar angle,  then we can also assume, without any loss of generality that
$R_{2}=0$. Vector $\bG_{0}$ is perpendicular to $\bR_{0}$, and its modulus is
fixed  by~\eqref{eq:c4e}. In \ed{the generic} case, $R_{1}R_{3}\neq 0$, and we can put:
 \begin{equation}
   \nonumber
   \bG_{0}=a_{1}A[0,1,0]^{T} +a_{2}\frac{A}{R_{0}}[R_{3},0,-R_{1}]^{T}, 
 \end{equation}
where $a_{1}$ and $a_{2}$ are arbitrary \ed{real} numbers satisfying the
following relation:
\begin{equation}
  \nonumber
  a_{1}^{2}+a_{2}^{2}=\frac{L^{2}}{(R_{0}^{2}+A)^{2}}.
\end{equation}
For each choice of  $a_{1}$ and $a_{2}$, we have $\bG_{0}$, and then the
corresponding value of $\bP_{0}$ is given by~\eqref{eq:p0}. Thus, a solution to
equations~\eqref{eq:spoke_get_key}  gives us a one-parameter family of the
relative equilibria. 

\begin{figure}
	\begin{center}
		\includegraphics[width=0.44\textwidth]{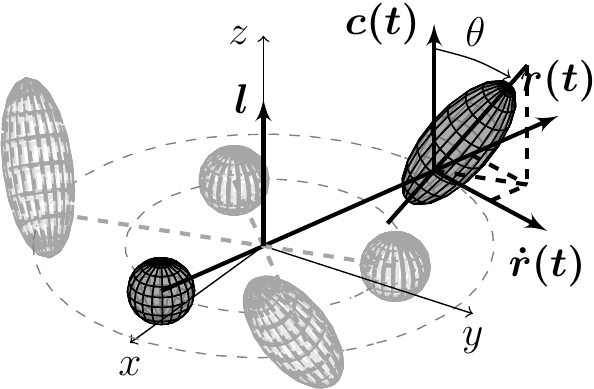}
	\end{center}
	\caption{\textit{The inclined co-planar precession} in the inertial reference
frame. Shaded meshes are for an illustration of the attitude and relative positions of the rigid body $\scB$ and the point
$\scP$ at different moments during the motion of the system.}
	\label{fig:fig4}
\end{figure}

From the above formulae it immediately follows that  point $\scP$,  and the mass
centre of the body $\scB$ move in one plane. In fact, we have:
\begin{equation}
  \vl\cdot\vr(t)
  =\pairing{\bG_{0}}{\bR_{0}}=0.
  \nonumber
\end{equation}
Moreover, as in the first case of the cylindrical precession, we can determine
the orientation of the symmetry axis using \eqref{orb}. We  have:
\begin{equation}
		\begin{aligned}
	\vr(t)\cdot\vz(t)
        &=\pairing{\bR_{0}}{ \bZ}=R_{3},\\
	\dot\vr(t)\cdot\vz(t)& =\pairing{\bP_{0}}{\bZ}=P_{3}=-a_{1}R_{1}, \\
\vc(t)\cdot\vz(t)&= \pairing{ \bR_{0}\times(
  \widetilde\bOmega\times\bR_{0})}{\bZ}= -a_{2}R_{0}R_{1}.
		\end{aligned}
		\nonumber
	\end{equation}
Thus, in general, the symmetry axis of the rigid body is inclined with respect 
to any
axis of the orbital reference frame. We shall call this class of 
relative equilibria as {\em
the inclined, co-planar regular precession}, and its geometry is illustrated in
Fig.~\ref{fig:fig4}.

\subsection{Non-locally central case \ed{(conic precession)}}
\label{s:conic}

As it has been already established, the non-locally central equilibria are
characterized by  $P_3=R_2\Omega_1-R_1\Omega_2=0$. It appears that in such 
\ed{an instance}, instead of using system ~\eqref{eq:sys_fullnoP}, it is better to analyse
\ed{equations} in which vector $\bP$ is not eliminated. Hence, we consider the
following \ed{system of equations}:
\begin{subequations}
	\begin{equation}
		\widetilde{\bOmega}\times\bR_0 = \bP_0,
		\label{seq:sys_full1}
	\end{equation}
	\begin{equation}
		\widetilde{\bOmega}\times\bP_0 = -\pder{U}{\bR_0},
		\label{seq:sys_full2}
	\end{equation}
	\begin{equation}
		\widetilde{\bOmega}\times\bG_0 = \bR_0\times\pder{U}{\bR_0},
		\label{seq:sys_full3}
	\end{equation}
	\begin{equation}
		\left( \bR_0\times\bP_0+\bG_0 \right) = L^2.
		\label{seq:sys_full4}
	\end{equation}
	\nonumber
\end{subequations}
Our aim is to find all solutions of this system which are not locally central.
Thus, we look for solutions for which  none of  variables  $\rho$, $R_3$,
$\widetilde{\Omega}_3$ vanishes. Hence,  we can express  $U_1$ and $U_2$ in terms
of $U_{\rho}$, i.e., 
\[
U_{i}=\frac{R_{i}}{\rho}U_{\rho}, \mtext{for} i=1,2.
\]
Since $P_3$ vanishes, from the first two components of
equations~\eqref{seq:sys_full2} it follows that
\begin{equation}
	P_1 = -R_2\frac{U_\rho}{\widetilde{\Omega}_3\rho},
	\mtext{and} 	P_2 =
	R_1\frac{U_\rho}{\widetilde{\Omega}_3\rho}. 
	\label{eq:sol3_P}
\end{equation}
Knowing  $P_1$ and $P_2$, from the first two components   of  equation
\eqref{seq:sys_full1}, we find:
\begin{equation}
	\Omega_i=
	R_i\frac{\rho\widetilde{\Omega}_3^2-U_\rho}{\rho
	R_3\widetilde{\Omega}_3}, \mtext{for} i=1,2.
	\label{eq:sol3_Omega}
\end{equation}
Substituting   \eqref{eq:sol3_P}, \eqref{eq:sol3_Omega} into the third
component of equation \eqref{seq:sys_full2}, we obtain:
\begin{equation}
	\left( \rho U_\rho+R_3U_3 \right)\widetilde{\Omega}_3^2 = U_\rho^2,
	\label{eq:Omega_quad}
\end{equation}
Now, let us consider equation~\eqref{seq:sys_full3}. It is  easy to see that 
the third component of this equation is fulfilled  identically.  Using
\eqref{eq:sol3_Omega}, the first two components can be rewritten  in
the  following form: 
\begin{equation}
	R_i\left( A\widetilde{\Omega}_3-C\Omega_3 \right)U_\rho U_3 = R_i\left( \rho U_\rho+R_3U_3 \right)\left( \rho U_3-R_3U_\rho \right)\widetilde{\Omega}_3, \mtext{\ed{for}} i=1,2.
	\nonumber
\end{equation}
As $R_1$ and $R_2$ do not vanish simultaneously, the above system is equivalent
to just one equation:
\begin{equation}
	\left( A\widetilde{\Omega}_3-C\Omega_3 \right)U_\rho U_3 = \left( \rho U_\rho+R_3U_3 \right)\left( \rho U_3-R_3U_\rho \right)\widetilde{\Omega}_3.
	\label{eq:Omega_tilde}
\end{equation}
Next, substituting expressions \eqref{eq:sol3_P} and ~\eqref{eq:sol3_Omega},
equation ~\eqref{seq:sys_full4} is rewritten as:
\begin{equation}
	L^2=\frac{\left[ \left( A+R_3^2 \right)U_3+\rho R_3U_\rho \right]^2}{\rho U_\rho+R_3U_3}+\left( \frac{\rho U_\rho}{\widetilde{\Omega}_3}+G_3 \right)^2,
	\label{eq:L_inter}
\end{equation}
Finally, we use~\eqref{eq:Omega_tilde} to eliminate $\widetilde{\Omega}_3$ from
equations~\eqref{eq:Omega_quad} and~\eqref{eq:L_inter}. Thus, eventually,
we obtain the final subsystem that makes it possible to determine the 
phase-space variables of the relative equilibrium, in \ed{quite a compact
form}:
%
%% this part has been shortened to avoid repetition and wide formulae
%
%\begin{equation}
%	\left.
%	\begin{aligned}
%		L^2U_3^2\left( \rho U_\rho+R_3U_3 \right) &= \left( U_\rho^2+U_3^2 \right)\left[ \left( A+R_3^2 \right)U_3+\rho R_3 U_\rho \right]^2,\\
%		U_3^2C^2\Omega_3^2\left( \rho U_\rho+R_3U_3 \right) &= \left[ AU_\rho U_3+\left( \rho U_\rho+R_3U_3 \right)\left( R_3U_\rho-\rho U_3 \right)
%\right]^2.\\
%	\end{aligned}
%	\quad\right\}
%	\nonumber
%\end{equation}
%
%It is still advantageous to introduce a new variable:
%
%
%
%which is helpful to simplify these conditions. 
%Then the system can be rewritten in  even more compact form:
%
\begin{equation}
	\left.
	\begin{aligned}
		L^2U_3^2K &= \left( U_\rho^2+U_3^2 \right)\left[ AU_3+R_3K \right]^2,\\
		C^2\Omega_3^2U_3^2K &= \left[ AU_\rho U_3+K\left( R_3U_\rho-\rho U_3 \right) \right]^2,\\
	\end{aligned}
	\quad\right\}
	\label{eq:sol3_final_short}
\end{equation}
\ed{where we introduced a new variable:}
\begin{equation}
	K\equiv \rho U_\rho + R_3U_3.
	\nonumber
\end{equation}
A solution to the above conditions gives us $\bR_0$. The remaining equilibrium
variables  can be determined by  equations ~\eqref{eq:sol3_P},
~\eqref{eq:sol3_Omega},  and ~\eqref{eq:Omega_quad}, respectively.

For this solution we can find,  after tedious simplifications,  that
\begin{equation}
  \vl\cdot\vr(t) =\pairing{\bG_{0}}{\bR_{0}}=
\frac{U_\rho \left( AU_3+R_3K \right)}{U_3 K}\left( R_3U_\rho-\rho U_3
\right).
  \nonumber
\end{equation}
According with our assumptions,  $\rho\neq 0$,  and hence $U_\rho\neq 0$. Thus,
the right hand side of the above equation vanishes if  either $R_3U_\rho-\rho
U_3=0$,  or  $AU_3+R_3K=0$. In the first case, $\bR_{0}$ is  locally central
point which we excluded from our considerations. In  the second 
instance,  from 
equation~\eqref{eq:sol3_final_short}, we obtain that  $L=0$, so it is also
excluded from our considerations. We conclude that non-locally central
relative equilibrium is {\em non-Lagrangian} ones. 

\begin{figure}
	\begin{center}
		\includegraphics[width=0.44\textwidth]{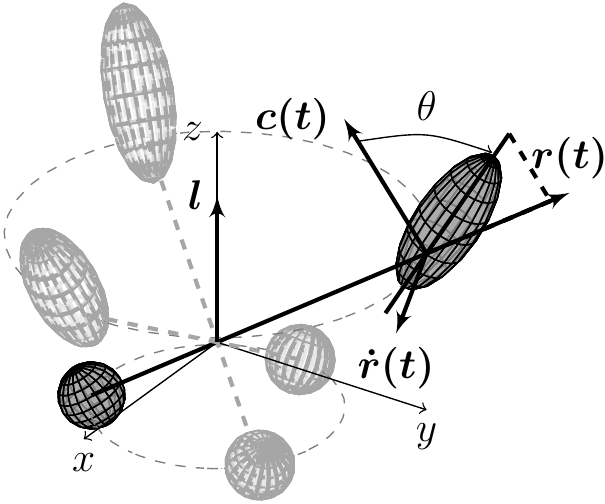}
	\end{center}
	\caption{\textit{Conic precession} in the inertial reference frame. Shaded bodies show the attitude of the rigid body $\scB$ and the point $\scP$ in other time points.}
	\label{fig:fig5}
\end{figure}

In order to determine the orientation of the symmetry axis in the orbital frame
which is specific for this type of relative equilibria, we can easily find that:
\begin{equation}
	\left.
	\begin{aligned}
		\vr(t)\cdot\vz(t) &= \pairing{\bR_{0}}{ \bZ}=R_{3},\\
		\dot\vr(t)\cdot\vz(t) &= 0,\\
		\vc(t)\cdot\vz(t) &= \pairing{ \bR_{0}\times(
		\widetilde\bOmega\times\bR_{0})}{\bZ}=\frac{\rho U_\rho}{\widetilde{\Omega}_3}.
	\end{aligned}
	\right.
	\nonumber
\end{equation}
Thus, in the orbital reference frame, the axis of symmetry always lies in the
plane formed by the relative position vector and the normal to the orbital
plane. We call this kind of the regular precession as {\em the conic precession}
from hereafter. Notice, that in the considered case, vector $\vc(t)$ is not
perpendicular to the orbital planes. Moreover, we have:
\begin{equation}
		\vl\cdot\vz(t) = \pairing{ \bR_{0}\times(
		\widetilde\bOmega\times\bR_{0})+\bG_0}{\bZ}=\frac{\rho
                U_\rho}{\widetilde{\Omega}_3}+C\Omega_3.
	\nonumber
\end{equation}
Thus, the axis of symmetry is inclined to the orbital plane. The geometry of
this solution is presented in Fig.~\ref{fig:fig5}.
Unlike the locally central cases, even for bodies with simple potentials, it
is difficult to decide whether equations~\eqref{eq:sol3_final_short} have a
solution.  In order to justify that the relative equilibria belonging to the conic
precession class \ed{may really} exist, we apply the following reasoning. If $\rho$,
as well as $\Omega_{3}$, tends to zero, then the conic precession tends to 
\ed{a particular} case of the inclined planar regular precession described in
Section~\ref{s:inclined_planar}.  In fact, if $\rho=\Omega_{3}=0$, then
$U_{\rho}=0$, so the second equation in~\eqref{eq:sol3_final_short} is satisfied
identically, and the first one  coincides  exactly  with the equation describing
corresponding case of the inclined co-planar solution.  Thus, we can take
this particular solution as the zero-th order  approximation (with respect to
$\Omega_{3}$) of the conic precession solution, and we apply the perturbative
approach to solve the system~\eqref{eq:sol3_final_short}.

Thus, the zero-th order solution is given through: 
\begin{equation}
	\rho=0,\quad R_3=R_3^*,
	\nonumber
\end{equation}
where $R_3^*$ is a solution of the following equation
\begin{equation}
	\left.	U_3 \right|_{\rho=0}=\frac{L^2}{\left( A+R_3^2 \right)^2}R_3.
	\nonumber
\end{equation}
To shorten the notation, we define:
\begin{equation}
	\left. U_3 \right|_{\rho=0,R_3=R_3^{*}}=F,\quad
	\left. U_{\rho\rho} \right|_{\rho=0,R_3=R_3^{*}}=V,\quad \left. U_{33} \right|_{\rho=0,R_3=R_3^{*}}=W.
	\nonumber
\end{equation}
Taking into account that
\[
	\left. U_\rho \right|_{\rho=0}=\left. U_{\rho 3}
        \right|_{\rho=0}=0,
\]
we can find that the Jacobian at the zeroth order solution has the form of:
\begin{equation}
	\bJ=
	\begin{bmatrix}
		0 & J_{12}\\
		J_{21} & 0\\
	\end{bmatrix},
	\nonumber
\end{equation}
where
\begin{subequations}
	\begin{equation}
		J_{12} = -\frac{L^6R_3^{*3}}{\left( A+R_3^{*2} \right)^7}\left[ \left( 3R_3^{*2}-A \right)L^2+W\left( A+R_3^{*2} \right)^3 \right],
		\label{seq:J_12}
	\end{equation}
	\begin{equation}
		J_{21} = -\frac{L^2R_{3}^{*}}{\left( A+R_3^{*2} \right)^4}\left[ V\left( A+R_3^{*2} \right)^3-L^2A \right],
		\label{seq:J_21}
	\end{equation}
\end{subequations}
The determinant of this Jacobian, given through:
\begin{equation}
	\det \bJ=-J_{12}J_{21},
	\nonumber
\end{equation}
does not vanish identically. Hence, by the implicit function theorem,
system~\eqref{eq:sol3_final_short} has a solution $\rho=\rho(\Omega_{3})$,
$R_{3}=R_{3}(\Omega_{3})$, such that $\rho(0)=0$, and $R_{3}(0)=R_{3}^{*}$.
Moreover, such solution is unique. 

We underline an important fact that, even if the body is symmetric with respect
to the equatorial plane, the described conic precession does exist. In other words,
splitting of the orbital planes can be induced not only by an asymmetric mass
distribution, it can be also induced by a proper, particular rotation of the rigid body.
\ed{Remarkably, the first type of solutions has been known since 
\cite{1979AZh....56..881A}, who constructed them for a class of non-symmetric
potentials and further analysed their consequences in the motion model of 
the Earth-Moon system \citep{Barkin1980}.}

\section{A comparison with the work of Kinoshita}
Our analysis presented in the previous sections lead to  somewhat different
conclusions than those ones in \citep{1970PASJ...22..383K}, and in this section we
give an explanation and overview of these discrepancies. 
Kinoshita found three classes of stationary solutions:
\begin{enumerate}
	\item {\em the \ed{``}float\ed{''}} case,  when the axis of symmetry of the rigid body is always
perpendicular to the orbital plane,
	\item {\em the \ed{``}spoke\ed{''}} case, when   the axis of
symmetry lies along the relative position vector,
	\item {\em the \ed{``}arrow\ed{''}} case, when 
the axis of symmetry lies in the plane formed by tangential and normal
to the orbital plane vectors.
\end{enumerate}
Now we may show that all these solutions are, in our terminology,
locally central and Lagrangian, i.e., the point mass $\scP$, and the mass centre
of the rigid body $\scB$ move in one plane.

\ed{Since for the description of the system there has been chosen the reference frame
fixed at mass centre of the point body $\scP$}, the potential of the body, 
besides the distance from the mass centre, 
depends also on the orientation of the rigid body. Following Kinoshita, we
denote a parameter, which describes the orientation, as $\nu$.  In this
designation, $\nu$ is an angle under which the point body $\scP$ is 
\ed{``seen''} from
the mass centre of the rigid one $\scB$. Note an important fact that the {\em
\ed{``}float\ed{``}} and {\em \ed{``}arrow\ed{''}} \ed{types} of motion correspond to a case when $U_\nu$
vanishes. Let us determine the physical meaning of that condition
in terms of \ed{our parameterization}. For that reason, we will use 
\ed{an expansion of the rigid body potential} through series in terms of the
Legendre polynomials, the same as  in \cite{1970PASJ...22..383K}. This
representation of the potential of $\scB$ is the following: 
\begin{equation}
	U = -\frac{1}{r}\sum_{k=0}^{\infty}J_k\left( \frac{a}{r} \right)^kP_k(\nu),
	\label{eq:compare_U}
\end{equation}
where $\nu$ is the cosine of the angle between the symmetry axis of the body and
the relative radius vector, and $P_{k}$ is the Legendre polynomial.   For the 
{\em ``float''}  and  the {\em ``arrow''} types of motion, $U_\nu$ vanishes. Hence, 
\begin{equation}
	U_\nu=\frac{1}{r}\sum_{k=0}^\infty J_k\left( \frac{a}{r} \right)^k P_k'(\nu)=0,
	\nonumber
\end{equation}
for a certain value of $\nu$.

In terms of variables used in our paper, the expansion of
\eqref{eq:compare_U} reads as follows:
\begin{equation}
	U = -\frac{1}{R_0}\sum_{k=0}^{\infty}J_k\left( \frac{a}{R_0}
        \right)^kP_k\left( \frac{R_3}{R_0} \right). 
	\nonumber
\end{equation}
Let us recall that a point is locally central if and only if  
\[
\rho U_3-R_3U_\rho=0. 
\]
Thus, we have:
\begin{equation}
	\rho U_3-R_3U_\rho= \frac{\rho}{R_0^4}\sum_{k=0}^\infty
	J_k\left( \frac{a}{R_0} \right)^k P_k'\left( \frac{R_3}{R_0}
	\right)=\frac{\rho}{R_0^4} U_{\nu}=0.
	\label{eq:compare_lc}
\end{equation}
If $U_{\nu}=0$ then it means that the considered solution is locally central. 
Condition $U_{\nu}=0$ includes all possible locally central cases except
of that
ones when $\rho$ vanishes. That is why \ed{the {\em ``arrow''} and the {\em ``float''} type
} relative equilibria found by
\cite{1970PASJ...22..383K}, under assumption that $U_\nu=0$, are locally
central. \ed{Moreover,
it is  easy to see , the {\em ``spoke''} solution  is also locally central}. In fact,
for this solution  we have $\rho=U_\rho=0$.  Thus, {\em all solutions  found by
Kinoshita are locally central}. 

The {\em ``float''}  equilibrium  coincides exactly with our first solution as in both of
these cases the axis of symmetry is always perpendicular to the orbital plane. 
The {\em ``arrow''}  and the {\em ``spoke''}  motions are just particular cases of our
second solution. Indeed, the second solution, when $\Omega_3$ vanishes, becomes
the {\em ``spoke''}  motion, i.e., the axis of symmetry lies along the relative position
vector\ed{; if we have $R_{3}=0$}, then it coincides with the
{\em ``arrow''} motion.  \ed{Quite surprisingly}, in the paper of Kinoshita, we did not find  any
other cases corresponding to the inclined planar solution.  It seems that
investigating the {\em ``arrow''} case, \ed{Kinoshita} assumed for simplicity, that the
body $\scB$ {\em is symmetric with respect to the equatorial plane}. Moreover, then he
used generally wrong implication saying that  $U_\nu=0$ means $\nu=0$. 
Let us notice that in our representation $\nu=0$ is equivalent to $R_{3}=0$.

To summarize \ed{this Section}, \cite{1970PASJ...22..383K}  did not find the
non-Lagrangian solutions because he implicitly assumed that \ed{one} can
always choose an inertial reference frame in such a way that the relative
orbit lies in its $(x,y)$-plane. \ed{In fact, this is equivalent to 
a priori assumption that all stationary solutions must be Lagrangian ones}.

\section{Conclusions}
In this paper, we performed global, geometric analysis of the stationary
motions in the unrestricted problem of a point mass and an axially symmetric
rigid body \ed{(the Kinoshita problem)}. Our aim was to determine all possible   relative equilibria in  this
problem. We have shown that three types of  stationary motions can be
distinguished. Two of them, \textit{the cylindrical precession} and \textit{the
inclined planar motion, are Lagrangian} and are characterized by
co-planar orbits of the mass centers of the bodies,  whereas the third type of
solutions, \textit{the conic precession},  is non-Lagrangian.

In \textit{the cylindrical precession} the axis of symmetry of the rigid body is
always perpendicular to the orbital plane. This type of stationary motion was
also found by \cite{1970PASJ...22..383K}, where it is called  the {\em ``float''}
motion.

In \textit{the inclined planar precession}, the axis of symmetry of the rigid
body is inclined to the orbit. Two special cases of such kind of motion  are
found by \cite{1970PASJ...22..383K}. In the first case the axis of symmetry lies
along the relative position vector, and in the second one lies   in the plane
formed by the  tangent and the normal to the orbit. In
\cite{1970PASJ...22..383K} these solutions are named as  the {\em ``spoke''} and the
{\em ``arrow''} motions,  respectively.

In \textit{the conic precession}, the axis of symmetry lies in the plane formed
by the relative position and the normal to the orbital plane vectors.  Moreover,
the point and the mass centre of the rigid body move in different parallel
planes.  This type of stationary motion is completely new in the problem
analyzed in this work, \ed{although, as we noted above, such solutions have been
constructed for specific non-symmetric potentials by
\cite{1979AZh....56..881A}.} We found them  \ed{thanks to a formulation of the
problem} through \ed{the minimal} set of assumptions, basically implying only
the form of the gravitational potential of the rigid body, besides \ed{the
analysis} of the problem in the very basic settings of the Newtonian dynamics.

We show that the determination of the stationary motions in the problem of
Kinoshita may be reduced to solving at most two non-linear, algebraic
equations.  Our results can be applied to an arbitrary axially symmetric body,
providing that an explicit form of its potential function is given.

In our forthcoming paper~\citep{2008..Veresh}, which is a direct continuation of
this work,  we perform the stability analysis of the relative equilibria found
here, and we apply our approach to study the dynamics of a few specific models
(i.e., choosing explicit form of the gravitational potential of the rigid
body). 

\section*{Acknowledgments}
We thank Antonio Elipe for the review of the manuscript.
This work has been supported by the European Commission
through the Astrodynamics Network under Marie Curie contract
MRTN-CT-2006-035151.

\bibliographystyle{mn2e}
%\bibliography{mathreva,books,main}
\newcommand{\noopsort}[1]{}

\label{lastpage}
\end{document}